\documentclass[aps,prx,twocolumn,superscriptaddress]{revtex4-2}
\usepackage[utf8]{inputenc}
\usepackage{graphicx}
\usepackage{amsmath}
\usepackage{amsfonts}
\usepackage{amssymb}
\usepackage{hyperref}
\usepackage{bm}

\graphicspath{{./}}

\begin{document}

\title{Resolution and Robustness Bounds for Reconstructive Spectrometers}

\author{Changyan Zhu}
\affiliation{Division of Physics and Applied Physics, School of Physical and Mathematical Sciences, Nanyang Technological University, Singapore 637371}
\affiliation{Centre for Disruptive Photonic Technologies, Nanyang Technological University, Singapore 637371}

\author{Hsuan Lo}
\affiliation{Division of Physics and Applied Physics, School of Physical and Mathematical Sciences, Nanyang Technological University, Singapore 637371}

\affiliation{Department of Physics, Massachusetts Institute of Technology, Cambridge, Massachusetts 02139}

\author{Jianbo Yu}
\affiliation{School of Electrical and Electronic Engineering, Nanyang Technological University, Singapore 637371}

\author{Qijie Wang}
\affiliation{School of Electrical and Electronic Engineering, Nanyang Technological University, Singapore 637371}

\author{Y.~D.~Chong}
\affiliation{Division of Physics and Applied Physics, School of Physical and Mathematical Sciences, Nanyang Technological University, Singapore 637371}
\affiliation{Centre for Disruptive Photonic Technologies, Nanyang Technological University, Singapore 637371}

\begin{abstract}
  Reconstructive spectrometers are an emerging class of devices that combine complex light scattering with inference.  Thus far, the physical determinants of their performance remain under-explored.  Within the regime of chaotic or diffusive scattering, the noise-induced error for spectral reconstruction is governed by Fisher information.  We use random matrix theory to derive a closed-form relation linking the variance bound to physical parameters: the spectral correlation length, mean transmittance, and the number of frequency and measurement channels.  This analysis reveals fundamental trade-offs between the physical parameters, and establishes the conditions for ``super-resolution'' below the limit set by the spectral correlation length.  Our theory is validated numerically using a random matrix model as well as full-wave simulations.  These results establish a physically-grounded framework for designing compact, performant and robust reconstructive spectrometers.
\end{abstract}

\maketitle
  
\textit{Introduction.}---Spectrometers are a diverse class of devices that decompose light into its spectral constituents for analysis.  They are traditionally divided into two types: dispersive spectrometers, which rely on geometric dispersion over long propagation paths, and interferometric spectrometers, which work by measuring optical path length differences \cite{davis2001fourier}.  Both approaches benefit from increased device size, and so face practical limits in miniaturization \cite{redding2013compact, momeni2009integrated, xia2011high}.  An alternative family of spectrometers, called reconstructive or computational spectrometers, has recently emerged \cite{yang2015miniature, yang2021miniaturization, zhang2021research}.  While their individual designs vary widely, a broad class of such devices work by encoding input light into a complex interference pattern, then using the measured speckle pattern to infer the spectrum via prior knowledge (e.g., calibration data).  Such schemes have been demonstrated in the visible, infrared, and terahertz regimes, using disordered waveguides \cite{redding2013compact}, multimode fibers \cite{redding2014using,liew2016broadband}, colloidal quantum dots \cite{bao2015colloidal}, and other platforms \cite{faraji2018compact, yang2019single, Lee2025Reconstructive}.

For conventional spectrometers, we have a sound understanding of how a device's performance depends on its physical characteristics; for instance, the pixel-limited spectral resolution of a dispersive spectrometer is known to scale as $1/L$, where $L$ is the optical path length.  For reconstructive spectrometers, however, the determinants of performance are less clear \cite{miller2019waves}.   A commonly-cited heuristic is the spectral correlation length $\Gamma_{\mathrm{corr}}$, the frequency scale over which output patterns decorrelate \cite{redding2013compact, yang2021miniaturization, yoon2022miniaturized, cai2024compact}.  It seems reasonable that a short $\Gamma_{\mathrm{corr}}$ is desirable, as it makes the patterns for neighboring frequencies more easily distinguishable, but there have been only limited efforts to study this systematically \cite{varytis2018design, ma2026inverse}; it is unclear if $\Gamma_{\mathrm{corr}}$ is the main constraint, or whether other physical properties such as transmittance, noise levels, etc., may be more relevant.  Moreover, the performance of a reconstructive spectrometer may depend on idiosyncratic features of the reconstruction algorithm, such as domain-specific priors, making it hard to draw general conclusions.

In this Letter, we develop a broadly-applicable figure of merit for the performance of a reconstructive spectrometer, and analyze its physical implications.  We abstract away most implementation details, retaining only the following assumptions: (i) the spectrum is inferred from light intensity measurements on a fixed set of output channels; (ii) each intensity measurement is subject to independent Gaussian noise; (iii) the spectrometer comprises a chaotic or diffusive scatterer with strongly-overlapping resonances (the Ericson regime) \cite{ericson1960fluctuations, lehmann1995chaotic}, and (iv) there are no other relevant priors about the spectrum (e.g., sparsity).  The number of measurement channels may be smaller than the number of frequency channels (as in highly miniaturized devices), making the inference problem under-determined; however, the over-determined regime is covered by the same framework.  The theory of Fisher information \cite{frieden1998physics} then implies that the noise-induced reconstruction error is bounded below by $\sigma_\epsilon^2 \operatorname{Tr}[(\mathbf{A}^T \mathbf{A})^+]$, where $\sigma_\epsilon$ is the noise level, $\mathbf{A}$ is the normalized spectral transmission matrix, and $(\cdots)^{+}$ denotes the Moore--Penrose pseudoinverse.  Leveraging on empirically-grounded features of generic $\mathbf{A}$ matrices \cite{fyodorov2016random, lehmann1995chaotic, fyodorov1997statistics, popoff2010measuring, goetschy2013filtering, cao2015dielectric}, we derive the noise-induced reconstruction error and effective resolution in terms of $\Gamma_{\mathrm{corr}}$, the average transmittance, and the number of frequency and measurement channels.  In particular, we obtain the necessary conditions for a reconstructive spectrometer to achieve ``super-resolution'' below the scale set by $\Gamma_{\mathrm{corr}}$.

We subject the theory to two kinds of numerical tests, using a random matrix theory (RMT) model and full-wave simulations.  The theory closely matches numerical outcomes over a wide range of conditions within the assumed scattering regime, either using the pseudoinverse or a neural network for inference.  Super-resolution is found to be achievable in both models.  In the full-wave simulations, we show that combining our theory with the scaling laws for diffusive transport \cite{akkermans2007mesoscopic} yields a correct prediction of optimal device size without a brute-force search.  Finally, we discuss the possibility that the systems best able to overcome these reconstruction error bounds are those \textit{outside} the Ericson regime, with non-Lorentzian spectral correlations~\cite{ericson1960fluctuations, lehmann1995chaotic}.

\begin{figure}
\includegraphics[width=0.92\columnwidth]{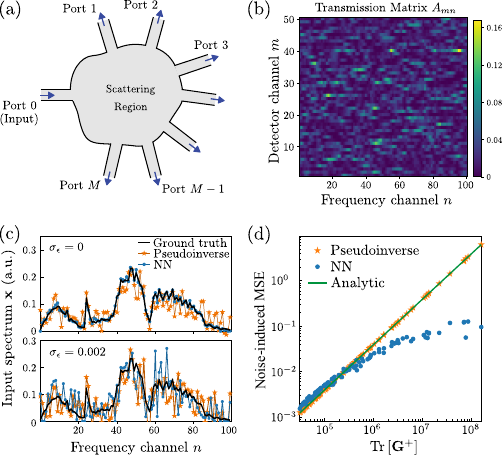}
\caption{(a) Schematic of a reconstructive spectrometer with $M$ output ports.  (b) Contents of a typical transmission matrix $\mathbf{A}$, of size $M = 50$ and $N = 100$, generated from the RMT model \cite{fyodorov2016random, SM}.  Each of the $N$ columns contains the output intensities at a given frequency.  (c) Simulated spectral reconstruction under zero noise (upper panel) and noise level $\sigma_\epsilon = 0.002$ (lower panel).  The ground truth (black line) is reconstructed via the transmission matrix from (b) using a pseudoinverse (orange stars), or a neural network (NN) trained on spectrum-intensity pairs (blue circles).  (d) Noise-induced mean squared error (MSE) versus $\operatorname{Tr}[\mathbf{G}^{+}]$, for an ensemble of $\mathbf{A}$ matrices with $\sigma_\epsilon = 0.002$.  The samples are generated by varying the RMT model's coupling strength $\gamma \in [0.002, 0.01]$, keeping all other parameters constant \cite{SM}.}
\label{fig:reconstruction}
\end{figure}

\textit{Spectral Reconstruction Model.}---A schematic of a reconstructive spectrometer is shown in Fig.~\ref{fig:reconstruction}(a).  Light enters an input port (0), undergoes linear scattering, and partly exits through a set of output ports $1$ to $M$.  The system's scattering matrix, $\mathbf{S}(\omega) \in \mathbb{C}^{(M+1)\times(M+1)}$, varies with frequency $\omega$.  We sample $N$ equally-spaced frequencies $\omega_1, \dots, \omega_N$.  The $M\times N$ spectral transmission matrix $\mathbf{A}$, defined by $A_{mn} = |S_{m0}(\omega_n)|^2 \in \mathbb{R}^+_0$, gives the intensity in each port and frequency channel.  In practice, $\mathbf{A}$ is obtained directly from calibration measurements (or simulations), with no additional normalization.  Fig.~\ref{fig:reconstruction}(b) shows a visualization of a typical $\mathbf{A}$ matrix.

For an input with power spectrum $\mathbf{x} \in (\mathbb{R}_0^+)^N$, the output intensities are $\mathbf{y} \in (\mathbb{R}_0^+)^M$.  We assume the measurements are subject to additive Gaussian noise,
\begin{equation}
  \mathbf{y} = \mathbf{A}\mathbf{x}+\boldsymbol{\epsilon},\quad
  \boldsymbol{\epsilon}
  \sim \mathcal{N}(\mathbf{0},\,\sigma_\epsilon^2\,\mathbf{1}_M),
  \label{eq:model_fisher}
\end{equation}
where $\sigma_\epsilon$ is the noise level.  (In the Supplemental Materials, we generalize the theory to Poissonian shot noise, and find no qualitative change in the predictions \cite{SM}.)  Thus, the average transmittance-to-noise ratio for the detection channels is $\mathcal{R}_0 \equiv \langle A_{mn}\rangle / \sigma_\epsilon$.  Given $\mathbf{y}$, we aim to infer $\mathbf{x}$, which reduces to inverting $\mathbf{A}$ if $\sigma_\epsilon = 0$ and $M = N$.  From the log-likelihood
$\log p(\mathbf{y}|\mathbf{x}) =-(M/2) \log(2\pi\sigma_\epsilon^2)
- (2\sigma_\epsilon^2)^{-1} \|\mathbf{y}-\mathbf{A}\mathbf{x}\|_2^2$, we obtain the Fisher matrix \cite{frieden1998physics}
\begin{equation}
  \mathbb{E}\big[\nabla_{\mathbf{x}}\log p\,\nabla_{\mathbf{x}}\log p^{\!\top}\big]
  = \sigma_\epsilon^{-2}\, \mathbf{G},
  \label{eq:fisher}
\end{equation}
where $\mathbf{G} = \mathbf{A}^T \mathbf{A}$.  For any unbiased estimator $\hat{\mathbf{x}}$, the covariance obeys the Cram\'er--Rao bound $\mathrm{Cov}(\hat{\mathbf{x}}) \succeq \sigma_\epsilon^{2}\,\mathbf{G}^{+}$, where $\mathbf{G}^{+} \equiv \mathbf{G}^T (\mathbf{G}\mathbf{G}^T)^{-1}$ for $M<N$, and $\mathbf{G}^{+} \equiv (\mathbf{G}^T \mathbf{G})^{-1}\mathbf{G}^T$ for $M\geq N$ \cite{kay1993fundamentals, ober2003calculation}.  The bound is saturated by the least squares (or maximum-likelihood) estimator if $\mathbf{A}$ has full column rank.  The trace gives a scalar bound for the mean squared error (MSE); if $\hat{\mathbf{x}}$ is biased, the MSE decomposes into an estimator variance and squared bias,
\begin{equation}
  \mathbb{E} \, \big\|\hat{\mathbf{x}}-\mathbf{x}\big\|_{2}^{2}
  \;\,\ge\;
  \sigma_\epsilon^{2}\,\mathrm{Tr}(\mathbf{G}^{+})
  \,+\, \big\| \, \mathbb{E}[\hat{\mathbf{x}}]-\mathbf{x} \, \big\|_{2}^{2}.
  \label{eq:mse-decomposition}
\end{equation}


To test this, we simulate spectral reconstruction with $\mathbf{A}$ matrices derived from random $\mathbf{S}$ matrices generated by the physically-motivated Mahaux--Weidenm\"uller model \cite{fyodorov2016random}, and $\mathbf{x}$'s derived from real spectra in the HITRAN dataset \cite{rothman1987hitran} (see Supplemental Materials, Sec.~S1 \cite{SM}).  In Fig.~\ref{fig:reconstruction}(c), the upper panel shows a typical outcome with $M = 50$ ports, $N = 100$ frequency channels, and zero noise ($\sigma_\epsilon = 0$).  Compared to the ground truth $\mathbf{x}$ (black line), the pseudoinverse estimator $\hat{\mathbf{x}} = \mathbf{A}^+ \mathbf{y}$ is fairly accurate (orange stars).  For comparison, we also train a dense neural network (NN) on a dataset of input-output pairs (see Supplemental Materials, Sec.~S1 \cite{SM}); this nonlinear method, which falls outside the scope of the above analysis, yields comparable or better results (blue circles).  Next, we add noise ($\sigma_{\epsilon} = 0.002$), and the revised results are shown in the lower panel of Fig.~\ref{fig:reconstruction}(c).  Both the pseudoinverse and NN estimates undergo degradation, which can be quantified by the ``noise-induced MSE,'' i.e., the difference in MSE with and without noise, which is equivalent to subtracting the bias term in \eqref{eq:mse-decomposition}.

Fig.~\ref{fig:reconstruction}(d) shows the noise-induced MSE for 100 samples of $\mathbf{A}$, generated by varying the coupling strength in the RMT model and checking 500 test spectra for each $\mathbf{A}$.  The predicted variation of the noise-induced MSE with $\mathrm{Tr}(\mathbf{G}^+)$ is the green line, which the pseudoinverse results follow closely.  For NN reconstruction, the noise-induced MSE initially follows the line, but falls below it at larger values of $\mathrm{Tr}(\mathbf{G}^+)$, which may be due to the NN exploiting prior information about the spectra to mitigate the effects of noise more effectively.  Overall, we conclude that the Cram\'er--Rao bound holds for both methods in the small $\mathrm{Tr}  (\mathbf{G}^+)$ regime, where noise is not dominant.  However, the pseudoinverse and NN can behave differently with respect to the bias part of the MSE, which is not plotted in Fig.~\ref{fig:reconstruction}(d).  The Cram\'er--Rao bound constrains only the noise-induced variance, not the \textit{total} MSE of Eq.~\eqref{eq:mse-decomposition}; specific algorithms, like the trained NN, can lower the bias term by learning the features of the spectra in the dataset.  We will focus on the weakly-noisy, prior-free regime, where the estimator variance dominates regardless of the reconstruction algorithm.  For a detailed comparison of the performance of the pseudoinverse and NN over a range of model parameters, see Supplemental Materials, Sec.~S2 \cite{SM}.


\textit{Physical variables.}---The Fisher information is an abstract quantity, so let us gain more insight into it by analyzing the $\mathbf{A}$ matrix.  Each column, $\mathbf{A}_n$, has the form
\begin{equation}
  \mathbf{A}_n = \langle \mathbf{A}_n\rangle + \mathbf{a}_n,
\end{equation}
where $\mathbf{a} = [\mathbf{a}_1,\dots, \mathbf{a}_{N}]$ is the mean-subtracted transmission matrix.  Therefore,
\begin{align}
  \mathbf{A}^{T} \mathbf{A} = N\bm{\mu}^{T} \bm{\mu} + \mathbf{C},
  \;\;\mathrm{where}\;\left\{
  \begin{aligned}
    \bm{\mu} & \equiv \langle A_{mn} \rangle \mathbf{1}_M \\
    \mathbf{C} &\equiv  \mathbf{a}^{T} \mathbf{a} \in \mathbb{R}^{N \times N},
  \end{aligned}\right.
  \label{eq:Cdef}
\end{align}
since the cross terms $\langle \mathbf{A}_n\rangle^T \mathbf{a}_{n'}$ vanish.  By the Sherman--Morrison formula \cite{sherman1950adjustment}, the rank-1 term $N\bm{\mu}^{T}\bm{\mu}$ modifies the inverse matrix in the direction defined by $\bm{\mu}$. For large $N$, the trace of the inverse is dominated by the bulk spectrum of $\mathbf{C}$, so this shift is negligible, and
\begin{equation}
  \mathrm{Tr}\big[\mathbf{(\mathbf{A}^T \mathbf{A})}^{+}\big] = \mathrm{Tr}\left[(N\bm{\mu}^{T}\bm{\mu} + \mathbf{C})^{+}\right]
  \approx \mathrm{Tr}[\mathbf{C}^{+}].
\end{equation}
Each element of $\mathbf{C}$ is the inner product of two zero-mean vectors $\mathbf{a}_i$ and $\mathbf{a}_j$.  If we assume speckle statistics typical of chaotic scattering media (with intensities following a Rayleigh distribution, and the standard deviation of the transmission equaling the mean \cite{goodman2007speckle}), the correlations can be approximated as
\begin{equation}
  \label{eq:correlation_C}
  \mathrm{corr}(\mathbf{A}_i, \mathbf{A}_j) = \frac{\langle \mathbf{a}_i, \mathbf{a}_j \rangle}{\|\mathbf{a}_i\|\|\mathbf{a}_j\|} \approx \frac{M\,C_{ij}}{T_0^2},
\end{equation}
where $T_0 \equiv M \langle A_{mn} \rangle$ is the total transmittance from the input to all $M$ outputs, averaged over all frequencies.  (For details, see Supplemental Materials, Sec.~S3 \cite{SM}.)

\begin{figure}
    \includegraphics[width=0.92\columnwidth]{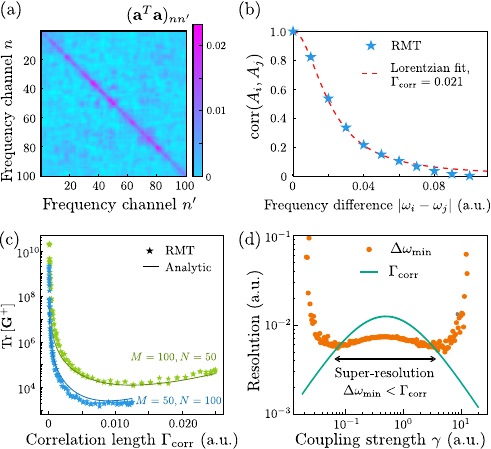}
    \caption{
      (a) Structure of the matrix $\mathbf{C} = \mathbf{a}^T\mathbf{a}$ for a typical scatterer, showing its near-Toeplitz form. The $\mathbf{A}$ matrix is generated by random matrix theory (RMT) using a Mahaux--Weidenm\"uller model \cite{fyodorov2016random}. (b) Calculated correlation function between columns of the $\mathbf{A}$ matrix from (a), showing a Lorentzian decay with frequency spacing.  The FWHM is denoted by $\Gamma_{\mathrm{corr}}$.  (c) Variation of $\operatorname{Tr}[\mathbf{G}^{+}]$ with $\Gamma_{\textrm{corr}}$, using an RMT ensemble in which the coupling parameter $\gamma$ varies from 0.005 to 0.1, with $N_{\mathrm{modes}} = 500$ internal cavity modes.  The results calculated directly from the $\mathbf{A}$ matrices (stars) agree with the analytic results based on $\Gamma_{\textrm{corr}}$ and the mean transmittance $T_0$ (solid curves), for both the under-determined and over-determined regimes. (d) Effective resolution $\Delta \omega_{\textrm{min}}$ and spectral correlation length $\Gamma_{\textrm{corr}}$ versus coupling strength $\gamma$.  Each data point is averaged over 200 samples, using $M=50$, $N=100$, and $N_{\mathrm{modes}} = 500$.  The effective resolution $\Delta \omega_{\textrm{min}}$ is calculated from Eq.~\eqref{eq:Emin_underdet} using $\delta^2_{\mathrm{th}} = 4\times10^{-4}$.  Over a range of $\gamma$, the device exhibits super-resolution ($\Delta \omega_{\textrm{min}} < \Gamma_{\textrm{corr}}$).}
	\label{fig.physical_properties}
\end{figure}

In many physical systems, spectral correlations decay with frequency separation (i.e., $|i-j|$ if the frequency channels are equally spaced) as Lorentzians \cite{ericson1960fluctuations, lehmann1995chaotic, fyodorov1997statistics, weidenmuller2009random, redding2013compact}.  If this holds, $\mathbf{C}$ is a \textit{near-Toeplitz} matrix, i.e., the expectation of $\mathcal{C} = \mathbb{E}[\mathbf{C}]$ takes a Toeplitz form \cite{gray2006toeplitz}:
\begin{equation}
  \mathcal{C}_{ij}
  = \frac{T_0^2}{M}\,
  \frac{a^{2}}{(i-j)^{2}+a^{2}},
  \qquad
  a\equiv\frac{\Gamma_{\rm corr}}{\Delta \omega}.
  \label{eq:toeplitz_lorentz}
\end{equation}
Here, $\Delta \omega$ is the sampling step in frequency space, and $\Gamma_{\rm corr}$ is the correlation length.  Each realization of $\mathbf{C}$ fluctuates around this mean form, as shown in Fig.~\ref{fig.physical_properties}(a); the spectral correlation is Lorentzian, as shown in Fig.~\ref{fig.physical_properties}(b).  The correlation length $\Gamma_{\rm corr}$ is the full width at half maximum (FWHM) of this spectral correlation curve.

\begin{figure*}
  \includegraphics[width=0.88\textwidth]{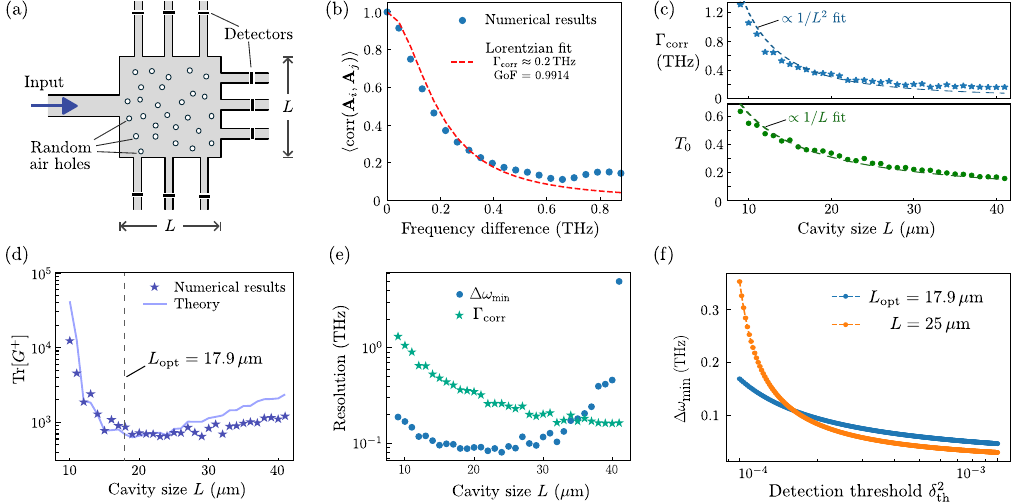}
  \caption{(a) Schematic of the on-chip reconstructive spectrometer with an $L\times L$ scattering region. (b) Spectral correlation function from FDTD simulations for $L=30\,\mu\textrm{m}$ (blue circles), fitted to a Lorentzian (red dashes) to extract $\Gamma_{\rm corr}$. (c) Scaling of $\Gamma_{\rm corr}$ (upper) and mean transmittance $T_0$ (lower) with cavity size $L$. Data points (averaged over 10 samples) agree well with the diffusive scaling laws $\Gamma_{\rm corr} \propto L^{-2}$ and $T_0 \propto L^{-1}$ (dashed lines). (d) Reconstruction error metric $\operatorname{Tr}[\mathbf{G}^{+}]$ versus $L$. The results obtained directly from the simulated $\mathbf{A}$ matrices (blue stars) are close to the theoretical predictions based on Eq.~\eqref{eq:inv-trace-mean} (blue line).  The minimum of the latter matches the optimal cavity size $L_{\mathrm{opt}}$ derived from the fits in (c) (vertical dashes).  (e) Plot of the effective resolution $\Delta\omega_{\textrm{min}}$ (orange circles) and $\Gamma_{\rm corr}$ (red circles) versus $L$, for detection threshold $\delta_{\mathrm{th}}^2 = 2 \times 10^{-4}$ and noise level $\sigma_{\epsilon} = 10^{-2}$.  (f) Effective resolution $\Delta\omega_{\textrm{min}}$ versus detection threshold $\delta_{\mathrm{th}}^2$, using two choices of cavity size: $L_{\mathrm{opt}} = 17.9\,\mu\textrm{m}$ (blue circles), which minimizes $\operatorname{Tr}[\mathbf{G}^{+}]$ as shown in (d); and $L=25\,\mu\textrm{m}$, which minimizes $\Delta\omega_{\textrm{min}}$ at $\delta_{\mathrm{th}}^2 = 2 \times 10^{-4}$, as shown in (e).  Here we again use $\sigma_{\epsilon} = 10^{-2}$.}
\label{fig.FDTD}
\end{figure*}

For large $N$, we can derive an asymptotic expression for $\mathrm{Tr}[\mathcal{C}^{+}]$ using the theory of Toeplitz matrices \cite{gray2006toeplitz, bottcher2006analysis}.  The key tool is Szeg\"{o}'s theorem \cite{hirschman1966strong}, which relates the trace of the inverse of a large Toeplitz matrix to an integral over its generating function $f(\theta)=\sum_{k=-\infty}^{\infty}C_k e^{ik\theta}$, where $C_k=C_{|k|}$.  First, consider the over-determined regime ($M\geq N$).  We simplify $f(\theta)$ via the identity
\begin{equation}
  \sum_{k=-\infty}^{\infty}
  \frac{e^{-ik\theta}}{k^{2}+a^{2}}
  =\frac{\pi}{a\sinh(\pi a)}
   \cosh\!\bigl[a\bigl(\pi-|\,\theta|\bigr)\bigr],
\end{equation}
and use Szeg\"o's theorem \cite{hirschman1966strong} to obtain
\begin{align}
  \frac{1}{N}\,{\rm Tr}\,\mathcal{C}^{+}
  &\overset{N\to\infty}{\longrightarrow} \; \frac{1}{2\pi}
    \int_{-\pi}^{\pi} \frac{d\theta}{f(\theta)}
    =\frac{M\,J(a)}{T_0^2},
    \label{eq:trace_Cinv} \\
  J(a) &\equiv \frac{2\sinh(\pi a)}{\pi^2 a^{2}}
  \tan^{-1}\!\Bigl[\tanh\bigl(\tfrac{\pi a}{2}\bigr)\Bigr].
  \label{eq:J_a}
\end{align}
Details of the derivation are given in the Supplemental Materials, Sec.~S3 \cite{SM}.  Hence, we find
\begin{align}
  \mathbb{E}\left[\,\operatorname{Tr}(\mathbf{G}^{+})\,\right]
  &= \frac{M}{|M-N|}{\rm Tr}\,\mathcal{C}^{+}
  = \frac{M^2 N}{|M-N|} \frac{J(a)}{T_0^{2}}.
  \label{eq:inv-trace-mean}
\end{align}
The $M/(M-N)$ factor arises from the Marchenko--Pastur distribution governing the spectrum of Wishart matrices \cite{tulino2004random}\footnote{Strictly speaking, the scattering matrices in the Mahaux--Weidenmüller model are not Wishart, but they approximately satisfy Wishart statistics in the chaotic regime with strong mode overlap.}.  (When $M \to N$, the ensemble average diverges due to ill-conditioning, but the noise MSE $\sigma_\epsilon^{2}\operatorname{Tr}[\mathbf{G}^{+}]$ remains finite; see Supplemental Materials, Fig.~S1(b) \cite{SM}.)   The normalized correlation length $a$
enters via $J(a)$, which is monotonically increasing.  For uncorrelated channels ($a < 1$), this has a minimum $J(0) \rightarrow 1$; for $a \gtrsim 1$, its asymptote is $J(a) \approx e^{\pi a}/(4\pi a^2)$; see Supplemental Materials, Sec.~S5 \cite{SM}.

We can translate the result into an effective spectral resolution $\Delta \omega_{\min}$, defined as the minimum frequency spacing to keep the reconstruction MSE below a threshold $\delta_{\text{th}}^2$, analogous to the Rayleigh criterion in classical optics.  For instance, when two spectral lines are separated by less than $\Delta \omega_{\min}$, they cannot be reliably distinguished; see Supplemental Materials, Sec.~S6 \cite{SM}.  In the correlated-channel regime ($a \gtrsim 1$), we find
\begin{align}
  \Delta \omega_{\min} &\approx \frac{\pi \Gamma_{\text{corr}}}{\ln \mathcal{R}^2}, \\
  \mathcal{R}
  &\approx
  \delta_{\text{th}} \, \sqrt{\frac{|M-N|}{N}} \, \mathcal{R}_0,
  \label{eq:Emin_overdet}
\end{align}
Eq.~\eqref{eq:Emin_overdet} is an effective signal-to-noise ratio for the device, expressed via the resources used to overcome noise in the system: the signal-to-noise ratio for a single channel ($\mathcal{R}_0$), the number of measurement channels $M$ relative to the $N$ unknowns, and the tolerance $\delta_{\mathrm{th}}$.  Notably, if $\mathcal{R}$ is sufficiently large, ``super-resolution'' below the scale of $\Gamma_{\text{corr}}$ is possible. 

In the under-determined case ($M<N$), $\mathbf{G}$ is not full rank, but the rank deficiency can be compensated for by replacing $J(a)$ with
\begin{equation}
  \tilde{J}(a) =  \frac{2\sinh\left(\sqrt{2}\frac{M}{N}\pi a\right)}{\pi^2 a^{2}}
    \tan^{-1}\!\left[\tanh\left(\frac{\pi a}{2}\right)\right].
    \label{eq:J_a_underdetermined}
\end{equation}
This function is derived from Fej\'er-window spectral smoothing and calibrated against RMT ensembles (see Supplemental Materials, Sec.~S4 \cite{SM}).  Eq.~\eqref{eq:inv-trace-mean} is thereupon replaced with
\begin{equation}  
  \mathbb{E}\left[\,\operatorname{Tr}(\mathbf{G}^{+})\,\right]
  \approx \frac{M^2 N}{|M-N|} \frac{\tilde{J}(a)}{T_0^{2}},
  \label{eq:inv-trace-mean-underdetermined}
\end{equation}
and the corresponding resolution in the $a \gg 1$ limit is
\begin{equation}
  \Delta \omega_{\min} \approx \frac{\sqrt{2}M}{N}
  \frac{\pi \Gamma_{\text{corr}}}{\ln \mathcal{R}^2}.
  \label{eq:Emin_underdet}
\end{equation}  
This result implies that increasing $N$ beyond $M$ can improve the resolution via the $M/N$ prefactor, provided $\mathcal{R}$ remains sufficiently large.  For excessively large $N$, however, $\mathcal{R}$ will eventually be driven below unity [Eq.~\eqref{eq:Emin_overdet}], at which point the resolution formula breaks down and spectral reconstruction becomes unfeasible.


To test these results, we generate RMT samples with different coupling strengths $\gamma$, and observe how $\operatorname{Tr}[\mathbf{G}^{+}]$ varies with $a$.  As shown in Fig.~\ref{fig.physical_properties}(c), the numerical results are very close to the predictions in both over-determined and under-determined regimes.  Contrary to the intuition that a short correlation length must be desirable \cite{yang2021miniaturization, yoon2022miniaturized, cai2024compact}, the reconstruction MSE varies non-monotonically with $a$ (since both $a$ and $T_0$ vary with $\gamma$).  In Fig.~\ref{fig.physical_properties}(d), we plot $\Delta \omega_{\min}$ versus $\gamma$ for $M < N$, and find that the super-resolution regime $\Delta \omega_{\min} < \Gamma_{\mathrm{corr}}$ is accessible in an interval of $\gamma$.


\textit{Full-wave simulations.}---We now demonstrate that our theory can provide quantitative guidelines for designing realistic devices. We perform full-wave finite-difference time-domain (FDTD) simulations of the two-dimensional structure in Fig.~\ref{fig.FDTD}(a), similar to the design in Ref.~\cite{redding2013compact}. The device comprises a square dielectric cavity of size $L \times L$, patterned with circular air holes (diameter $d_0 = 0.4\,\mu\textrm{m}$) in a Poisson-disk distribution with mean spacing $d \approx 1.8 d_0$. A transverse-electric Gaussian pulse (78.89--83.28 THz) is injected, and the scattered light is collected by $M = 9$ output waveguides. The spectral range is discretized into $N = 100$ channels to extract the transmission matrix $\mathbf{A}$.  Fig.~\ref{fig.FDTD}(b) shows the spectral correlation function for samples of size $L =30\,\mu \text{m}$, which exhibits a Lorentzian lineshape, consistent with our assumptions.  We now systematically generate ensembles of $\mathbf{A}$ by varying $L$ (which serves as a proxy for the optical path length) from 10 to 40~$\mu$m, with all other parameters fixed. 

Figure~\ref{fig.FDTD}(c) shows how the scattering properties depend on $L$.  For diffusive transport, $T_0 \propto \ell/L$ and $\Gamma_{\mathrm{corr}} \propto D/L^{2}$, where $\ell$ is the transport mean free path and $D$ is the diffusion constant \cite{akkermans2007mesoscopic}.  Hence, we can deduce the optimal cavity size without a brute-force search.  We aim to minimize the reconstruction error $\mathcal{E}[a(L)] \propto \tilde{J}[a(L)]/a(L)$; solving $d\mathcal{E}/da = 0$ numerically for $M=9$ and $N=100$, using Eq.~\eqref{eq:J_a_underdetermined}, yields $a_{\mathrm{opt}} \approx 7.46$.  Using the fitted diffusion constant $D \approx 120 \, \mu \textrm{m}^2/\mathrm{THz}$ and frequency channel spacing $\Delta \omega = 0.0439$ THz, we obtain an optimal cavity size $L_{\mathrm{opt}} = \sqrt{D/(a_{\mathrm{opt}} \cdot \Delta \omega)} \approx 17.9\,\mu\text{m}$, matching the observed minimum of $\operatorname{Tr}[\mathbf{G}^{+}]$ as shown in Fig.~\ref{fig.FDTD}(d).  This also illustrates a key trade-off highlighted by the theory: for $L < L_{\mathrm{opt}}$, performance is limited by excessive spectral correlations, whereas for $L > L_{\mathrm{opt}}$ it is limited by low optical throughput.

In Fig.~\ref{fig.FDTD}(e), we plot the effective resolution $\Delta \omega_{\textrm{min}}$ and $\Gamma_{\text{corr}}$ versus $L$, using noise level $\sigma_{\epsilon} = 10^{-2}$ (corresponding to $1 \lesssim \mathcal{R}_0 \lesssim 7$).  It can be seen that the spectrometer operates in the super-resolution regime below a certain cavity size, and the previously-obtained value of $L_{\mathrm{opt}}$ lies in this regime. Note that our previously-derived $L_{\mathrm{opt}}$ is near, but not equal to, the minimum of $\Delta \omega_{\textrm{min}}$ in this plot.  These are two related but distinct design targets one must choose between: the former maximizes the overall fidelity (by minimizing the noise-induced MSE), while the latter optimizes the effective frequency resolution at a given noise level and detection threshold.  To clarify this, Fig.~\ref{fig.FDTD}(f) shows the effective resolution $\Delta \omega_{\textrm{min}}$ versus detection threshold $\delta_{\mathrm{th}}^2$ for two cavity sizes: (i) $L = L_{\mathrm{opt}}$, and (ii) $L=25\,\mu\textrm{m}$, where the resolution curve in Fig.~\ref{fig.FDTD}(e) reaches its minimum.  The latter indeed yields a finer spectral resolution when operating at $\delta_{\mathrm{th}}^2 = 2\times10^{-4}$, the detection threshold used in Fig.~\ref{fig.FDTD}(e); for smaller $\delta_{\mathrm{th}}^2$, it gives a \textit{worse} (i.e., larger) resolution.


\textit{Conclusions.}---The performance of a reconstructive speckle spectrometer is largely determined by the variance bound $\operatorname{Tr}[\mathbf{G}^{+}]$, derived from its spectral transmission matrix.  When noise is present but non-dominant, this quantity governs the noise-induced error for not only pseudoinverse-based reconstruction, but also inference using a neural network.  We have derived $\operatorname{Tr}[\mathbf{G}^{+}]$, and hence the effective resolution, in terms of the spectral correlation length $\Gamma_{\mathrm{corr}}$, mean transmittance $T_0$, the number of intensity measurement channels, etc.   The theory yields successful predictions for both RMT models and a full-wave simulation of an on-chip device.  Our tests confirm that $\Gamma_{\mathrm{corr}}$ is not the sole determinant of performance; $T_0$ also plays a role and can lead to a trade-off.

A key assumption is that the scatterer exhibits conventional speckle statistics with Lorentzian spectral correlations \cite{ericson1960fluctuations, lehmann1995chaotic, fyodorov1997statistics, weidenmuller2009random}.  This holds for almost all scattering-based reconstructive spectrometers shown to date, including those based on photonic chips \cite{redding2013compact}, multimode fibers \cite{redding2014using}, colloids \cite{bao2015colloidal}, and nanostructures \cite{yang2019single, Lee2025Reconstructive}, but it is not universal.  In the Supplemental Materials \cite{SM}, we show that when a scatterer is expressly inverse-designed \cite{Molesky2018, Kroker2024, ma2026inverse} to minimize $\operatorname{Tr}[\mathbf{G}^{+}]$ \cite{yu2025wavelength}, the generated structure can have highly non-Lorentzian correlations while surpassing the predicted performance bounds.  This hints that non-Lorentzian speckle statistics, previously studied in the context of fundamental wave physics \cite{Lakshminarayan2008, chabanov2000statistical}, may have important applications.  In future, it will be desirable to find additional theoretical techniques to model such systems, such as the extremal statistics of RMT models \cite{Lakshminarayan2008}, which may allow us to characterize the ultimate limits of reconstructive spectrometry.

Other extensions of this framework are also interesting to pursue.  The two key theoretical ingredients, Fisher information and a spectral correlation law, can be applied to other schemes for reconstructing a spectrum (or other parameter distribution) from multiplexed measurements on complex scatterers, such as spectral or parametric sensing based on chaotic microwave cavities or acoustic reverberation chambers.  The role of priors may also be further interrogated; when there are important priors (e.g., spectral sparsity, or priors learned from training data), similar results may be obtainable by adopting Bayesian generalizations of the Cram\'er--Rao bound such as the van Trees inequality \cite{vantrees1968}.

We thank A.~Nussupbekov, B.~Zhu, F.~J.~N.~Jorgensen, and S.~G.~Johnson for helpful discussions.  This work was supported by the Singapore National Research Foundation (NRF) under the NRF Investigatorship NRF-NRFI08-2022-0001, and Competitive Research Program (CRP) Nos.~NRF-CRP23-2019-0005, NRF-CRP23-2019-0007, and NRF-CRP29-2022-0003.

\bibliography{spectrometer}

@article{fyodorov1997statistics,
  title={Statistics of resonance poles, phase shifts and time delays in quantum chaotic scattering: Random matrix approach for systems with broken time-reversal invariance},
  author={Fyodorov, Yan V and Sommers, Hans-J{\"u}rgen},
  journal={J. Math. Phys.},
  volume={38},
  number={4},
  pages={1918--1981},
  year={1997},
  publisher={American Institute of Physics}
}

@article{redding2013compact,
  title={Compact spectrometer based on a disordered photonic chip},
  author={Redding, Brandon and Liew, Seng Fatt and Sarma, Raktim and Cao, Hui},
  journal={Nat. Photon.},
  volume={7},
  number={9},
  pages={746--751},
  year={2013},
  publisher={Nature Publishing Group}
}

@article{yang2021miniaturization,
  title={Miniaturization of optical spectrometers},
  author={Yang, Zongyin and Albrow-Owen, Tom and Cai, Weiwei and Hasan, Tawfique},
  journal={Science},
  volume={371},
  number={6528},
  pages={eabe0722},
  year={2021},
  publisher={American Association for the Advancement of Science}
}

@article{lehmann1995chaotic,
  title={Chaotic scattering: the supersymmetry method for large number of channels},
  author={Lehmann, N and Saher, D and Sokolov, VV and Sommers, H-J},
  journal={Nucl. Phys. A},
  volume={582},
  number={1-2},
  pages={223--256},
  year={1995},
  publisher={Elsevier}
}

@inproceedings{fyodorov2016random,
  title={Random matrix theory of resonances: An overview},
  author={Fyodorov, Yan V},
  booktitle={2016 URSI International Symposium on Electromagnetic Theory (EMTS)},
  pages={666--669},
  year={2016},
  organization={IEEE}
}

@article{liew2016broadband,
  title={Broadband multimode fiber spectrometer},
  author={Liew, Seng Fatt and Redding, Brandon and Choma, Michael A and Tagare, Hemant D and Cao, Hui},
  journal={Opt. Lett.},
  volume={41},
  number={9},
  pages={2029--2032},
  year={2016},
  publisher={Optical Society of America}
}

@article{Lee2025Reconstructive,
  title   = {Reconstructive spectrometer using double-layer disordered metasurfaces},
  author  = {Lee, Dong-Gu and Song, Gookho and Lee, Chunghyung and Lee, Chanseok and Jang, Mooseok},
  journal = {Sci. Adv.},
  year    = {2025},
  volume  = {11},
  number  = {22},
  pages   = {eadv2376},
  doi     = {10.1126/sciadv.adv2376},
  url     = {https://www.science.org/doi/10.1126/sciadv.adv2376}
}

@article{yang2015miniature,
  title={Miniature spectrometer based on diffraction in a dispersive hole array},
  author={Yang, Tao and Xu, Cao and Ho, Ho-pui and Zhu, Yong-yuan and Hong, Xu-hao and Wang, Qian-jin and Chen, Yu-chao and Li, Xing-ao and Zhou, Xin-hui and Yi, Ming-dong and others},
  journal={Opt. Lett.},
  volume={40},
  number={13},
  pages={3217--3220},
  year={2015},
  publisher={Optical Society of America}
}

@article{zhang2021research,
  title={Research progress on on-chip Fourier transform spectrometer},
  author={Zhang, Lichao and Chen, Jiamin and Ma, Chaowei and Li, Wangzhe and Qi, Zhimei and Xue, Ning},
  journal={Laser Photon. Rev.},
  volume={15},
  number={9},
  pages={2100016},
  year={2021},
  publisher={Wiley Online Library}
}

@book{davis2001fourier,
  title={Fourier transform spectrometry},
  author={Davis, Sumner P and Abrams, Mark C and Brault, James W},
  year={2001},
  publisher={Academic press}
}

@article{cai2024compact,
  title={Compact angle-resolved metasurface spectrometer},
  author={Cai, Guiyi and Li, Yanhao and Zhang, Yao and Jiang, Xiong and Chen, Yimu and Qu, Geyang and Zhang, Xudong and Xiao, Shumin and Han, Jiecai and Yu, Shaohua and others},
  journal={Nat. Mat.},
  volume={23},
  number={1},
  pages={71--78},
  year={2024},
  publisher={Nature Publishing Group UK London}
}

@article{yoon2022miniaturized,
  title={Miniaturized spectrometers with a tunable van der Waals junction},
  author={Yoon, Hoon Hahn and Fernandez, Henry A and Nigmatulin, Fedor and Cai, Weiwei and Yang, Zongyin and Cui, Hanxiao and Ahmed, Faisal and Cui, Xiaoqi and Uddin, Md Gius and Minot, Ethan D and others},
  journal={Science},
  volume={378},
  number={6617},
  pages={296--299},
  year={2022},
  publisher={American Association for the Advancement of Science}
}

@inproceedings{redding2014using,
  title={Using a multimode fiber as a high-resolution, low-loss spectrometer},
  author={Redding, B and Cao, H},
  booktitle={Fiber Optic Sensors and Applications XI},
  volume={9098},
  pages={89--93},
  year={2014},
  organization={SPIE}
}

@article{momeni2009integrated,
  title={Integrated photonic crystal spectrometers for sensing applications},
  author={Momeni, Babak and Hosseini, Ehsan Shah and Askari, Murtaza and Soltani, Mohammad and Adibi, Ali},
  journal={Opt. Commun.},
  volume={282},
  number={15},
  pages={3168--3171},
  year={2009},
  publisher={Elsevier}
}

@article{xia2011high,
  title={High resolution on-chip spectroscopy based on miniaturized microdonut resonators},
  author={Xia, Zhixuan and Eftekhar, Ali Asghar and Soltani, Mohammad and Momeni, Babak and Li, Qing and Chamanzar, Maysamreza and Yegnanarayanan, Siva and Adibi, Ali},
  journal={Opt. Ex.},
  volume={19},
  number={13},
  pages={12356--12364},
  year={2011},
  publisher={Optical Society of America}
}

@article{varytis2018design,
  title={Design study of random spectrometers for applications at optical frequencies},
  author={Varytis, Paris and Huynh, Dan-Nha and Hartmann, Wladislaw and Pernice, Wolfram and Busch, Kurt},
  journal={Opt. Lett.},
  volume={43},
  number={13},
  pages={3180--3183},
  year={2018},
  publisher={Optical Society of America}
}

@article{gray2006toeplitz,
  title={Toeplitz and circulant matrices: A review},
  author={Gray, Robert M and others},
  journal={Foundations and Trends in Communications and Information Theory},
  volume={2},
  number={3},
  pages={155--239},
  year={2006},
  publisher={Now Publishers, Inc.}
}

@article{hirschman1966strong,
  title={The strong Szeg{\"o} limit theorem for Toeplitz determinants},
  author={Hirschman Jr, II},
  journal={American Journal of Mathematics},
  pages={577--614},
  year={1966},
  publisher={JSTOR}
}

@article{ober2003calculation,
  title={Calculation of the Fisher information matrix for multidimensional data sets},
  author={Ober, Raimund J and Zou, Qiyue and Lin, Zhiping},
  journal={IEEE transactions on signal processing},
  volume={51},
  number={10},
  pages={2679--2691},
  year={2003},
  publisher={IEEE}
}

@article{rothman1987hitran,
  title={The HITRAN database: 1986 edition},
  author={Rothman, Laurence S and Gamache, Robert R and Goldman, Aaron and Brown, Linda R and Toth, Robert A and Pickett, Herbert M and Poynter, Robert L and Flaud, J-M and Camy-Peyret, C and Barbe, A and others},
  journal={Applied optics},
  volume={26},
  number={19},
  pages={4058--4097},
  year={1987},
  publisher={Optical Society of America}
}

@book{frieden1998physics,
  title={Physics from Fisher information: a unification},
  author={Frieden, B Roy},
  year={1998},
  publisher={Cambridge University Press}
}

@misc{SM,
year={2025},
title = {See Supplementary Material},
note = {See online Supplemental Materials.}
}

@article{yu2025wavelength,
  title={Wavelength-scale noise-resistant on-chip spectrometer},
  author={Yu, Jianbo and others},
  journal={arXiv preprint arXiv:2509.22286},
  year={2025}
}

@article{sherman1950adjustment,
  title={Adjustment of an inverse matrix corresponding to a change in one element of a given matrix},
  author={Sherman, Jack and Morrison, Winifred J},
  journal={Ann. Math. Stat.},
  volume={21},
  number={1},
  pages={124--127},
  year={1950},
  publisher={JSTOR}
}

@article{bao2015colloidal,
  title={A colloidal quantum dot spectrometer},
  author={Bao, Jie and Bawendi, Moungi G},
  journal={Nature},
  volume={523},
  number={7558},
  pages={67--70},
  year={2015},
  publisher={Nature Publishing Group UK London}
}

@article{faraji2018compact,
  title={Compact folded metasurface spectrometer},
  author={Faraji-Dana, MohammadSadegh and Arbabi, Ehsan and Arbabi, Amir and Kamali, Seyedeh Mahsa and Kwon, Hyounghan and Faraon, Andrei},
  journal={Nat. Commun.},
  volume={9},
  number={1},
  pages={4196},
  year={2018},
  publisher={Nature Publishing Group UK London}
}

@article{yang2019single,
  title={Single-nanowire spectrometers},
  author={Yang, Zongyin and Albrow-Owen, Tom and Cui, Hanxiao and Alexander-Webber, Jack and Gu, Fuxing and Wang, Xiaomu and Wu, Tien-Chun and Zhuge, Minghua and Williams, Calum and Wang, Pan and others},
  journal={Science},
  volume={365},
  number={6457},
  pages={1017--1020},
  year={2019},
  publisher={American Association for the Advancement of Science}
}

@article{miller2019waves,
  title={Waves, modes, communications, and optics: a tutorial},
  author={Miller, David AB},
  journal={Adv. Opt. Photon.},
  volume={11},
  number={3},
  pages={679--825},
  year={2019},
  publisher={Optical Society of America}
}

@article{popoff2010measuring,
  title={Measuring the Transmission Matrix in Optics: An Approach to the Study and Control of Light Propagation in Disordered Media},
  author={Popoff, S{\'e}bastien M and Lerosey, Geoffroy and Carminati, R{\'e}mi and Fink, Mathias and Boccara, Albert Claude and Gigan, Sylvain},
  journal={Phys. Rev. Lett.},
  volume={104},
  number={10},
  pages={100601},
  year={2010},
  publisher={APS}
}

@article{goetschy2013filtering,
  title={Filtering random matrices: the effect of incomplete channel control in multiple scattering},
  author={Goetschy, A and Stone, AD},
  journal={Phys. Rev. Lett.},
  volume={111},
  number={6},
  pages={063901},
  year={2013},
  publisher={APS}
}

@article{cao2015dielectric,
  title={Dielectric microcavities: Model systems for wave chaos and non-Hermitian physics},
  author={Cao, Hui and Wiersig, Jan},
  journal={Rev. Mod. Phys.},
  volume={87},
  number={1},
  pages={61--111},
  year={2015},
  publisher={APS}
}

@book{kay1993fundamentals,
  title={Fundamentals of statistical signal processing: estimation theory},
  author={Kay, Steven M},
  year={1993},
  publisher={Prentice-Hall, Inc.}
}

@book{bottcher2006analysis,
  title={Analysis of Toeplitz operators},
  author={B{\"o}ttcher, Albrecht and Silbermann, Bernd},
  year={2006},
  publisher={Springer}
}

@article{ericson1960fluctuations,
  title={Fluctuations of nuclear cross sections in the" continuum" region},
  author={Ericson, Torleif},
  journal={Phys. Rev. Lett.},
  volume={5},
  number={9},
  pages={430},
  year={1960},
  publisher={APS}
}

@article{weidenmuller2009random,
  title={Random matrices and chaos in nuclear physics: Nuclear structure},
  author={Weidenm{\"u}ller, HA and Mitchell, GE},
  journal={Rev. Mod. Phys.},
  volume={81},
  number={2},
  pages={539--589},
  year={2009},
  publisher={APS}
}

@article{Molesky2018,
  title={Inverse design in nanophotonics},
  author={Molesky, Sean and Lin, Zin and Piggott, Alexander Y and Jin, Weiliang and Vuckovi{\'c}, Jelena and Rodriguez, Alejandro W.},
  journal={Nat. Photon.},
  volume={12},
  pages={659--670},
  year={2019},
  publisher={Nature Publishing Group}
}

@article{Kroker2024,
author = {Stefanie Kroker and St\'{e}phane Lanteri and Owen Miller and Jens Niegemann and Lora Ramunno},
journal = {J. Opt. Soc. Am. B},
number = {2},
pages = {IDP1--IDP2},
publisher = {Optica Publishing Group},
title = {Inverse design in photonics: introduction},
volume = {41},
month = {Feb},
year = {2024}
}

@book{goodman2007speckle,
  title={Speckle phenomena in optics: theory and applications},
  author={Goodman, Joseph W},
  year={2007},
  publisher={Roberts and Company Publishers}
}

@article{tulino2004random,
  title={Random matrix theory and wireless communications},
  author={Tulino, Antonia M and Verd{\'u}, Sergio and others},
  journal={Found. Trends Commun. Info. Theory},
  volume={1},
  number={1},
  pages={1--182},
  year={2004},
  publisher={Now Publishers, Inc.}
}

@article{Lakshminarayan2008,
  title = {Extreme Statistics of Complex Random and Quantum Chaotic States},
  author = {Lakshminarayan, Arul and Tomsovic, Steven and Bohigas, Oriol and Majumdar, Satya N.},
  journal = {Phys. Rev. Lett.},
  volume = {100},
  issue = {4},
  pages = {044103},
  numpages = {4},
  year = {2008},
  month = {Jan},
  publisher = {American Physical Society},
  doi = {10.1103/PhysRevLett.100.044103},
  url = {https://link.aps.org/doi/10.1103/PhysRevLett.100.044103}
}

@book{akkermans2007mesoscopic,
  title={Mesoscopic physics of electrons and photons},
  author={Akkermans, Eric and Montambaux, Gilles},
  year={2007},
  publisher={Cambridge university press}
}

@article{chabanov2000statistical,
  title={Statistical signatures of photon localization},
  author={Chabanov, AA and Stoytchev, M and Genack, AZ},
  journal={Nature},
  volume={404},
  number={6780},
  pages={850--853},
  year={2000},
  publisher={Nature Publishing Group UK London}
}

@book{vantrees1968,
  author = {Van Trees, Harry L.},
  title = {Detection, Estimation, and Modulation Theory, Part I},
  publisher = {Wiley},
  address = {New York},
  year = {1968}
}

@article{ma2026inverse,
  title={Inverse design for robust inference in integrated computational spectrometry},
  author={Ma, Wenchao and Pestourie, Rapha{\"e}l and Lin, Zin and Johnson, Steven G},
  journal={Nanophotonics},
  volume={15},
  number={7},
  pages={e70054},
  year={2026},
  publisher={Wiley Online Library}
}

\clearpage

\begin{widetext}
\makeatletter 
\renewcommand{\theequation}{S\arabic{equation}}
\makeatother
\setcounter{equation}{0}

\makeatletter 
\renewcommand{\thesection}{S\arabic{section}} 
\setcounter{section}{0}

\makeatletter 
\renewcommand{\thefigure}{S\@arabic\c@figure}
\makeatother
\setcounter{figure}{0}

\makeatletter 
\renewcommand{\thetable}{S\@arabic\c@table}
\makeatother
\setcounter{table}{0}

\begin{center}
  \textbf{Supplemental Materials for}\\
  \vskip 0.1in
  {\large ``Resolution and Robustness Bounds for Reconstructive Spectrometers''}\\
  \vskip 0.1in
\end{center}

\section{Dataset preparation and reconstruction algorithms}
\label{sec:sm_dataset}

\textit{Dataset preparation}---In our numerical simulations, the ground truth spectrum samples are drawn from a dataset constructed from the HITRAN (High-Resolution Transmission) spectroscopic database~\cite{rothman1987hitran}. From the database, we extract absorption spectra within the wavelength range 3600-$3800\,\textrm{nm}$ (78.9--83.3THz), selecting 64 experimental spectra corresponding to different isotopologues from 55 distinct molecules. Each spectrum is normalized to unit peak intensity and resampled onto a fixed frequency grid.  We then generate synthetic spectra emulating mixtures that may be encountered in sensing applications, by randomly combining 2--5 distinct experimental spectra per sample. This procedure yields a dataset of $10^4$ samples, of which 9000 are used for training the neural networks (see below),  and 1000 for testing.



\textit{Reconstruction algorithms}---During our numerical tests, we evaluate two spectral reconstruction algorithms:

\begin{enumerate}
\item The Moore-Penrose pseudoinverse.  The estimator is $\hat{x} = A^{+} y$, where $A^{+} = A^T (A A^T)^{-1}$.  This linear reconstruction scheme serves as the baseline for the theoretical analysis based on Fisher information and the Cram\'er--Rao bound.

\item A fully-connected neural network (NN), implemented using the PyTorch framework.  The NN comprises three hidden layers of 500 neurons each. Each layer is followed by a ReLU activation, dropout regularization, and batch normalization to enhance training stability and mitigate overfitting.  Training is done using the Adam optimizer with a learning rate of $1\times 10^{-3}$ for 200 epochs; the loss function is the mean squared error (MSE) between the predicted and true spectra.  Hyperparameters are optimized via grid search to achieve the best performance on a validation set.
\end{enumerate}

\section{RMT simulations}
\label{sec:RMT}

To model an ensemble of generic scatterers, we use a random matrix theory (RMT) model based on the  Mahaux--Weidenmüller formalism \cite{fyodorov2016random}.  A cavity is described by a $P\times P$ Hamiltonian $\mathbf{H}$ sampled from the Gaussian unitary ensemble (GUE), and its coupling to $M+1$ external channels is characterized by a $P\times(M+1)$ random matrix $\mathbf{W}$, where
\begin{equation}
  W_{pm} \sim \mathcal{N}(0, \gamma^2), \;\;\; \gamma \in \mathbf{R}^+.
  \label{Wdef}
\end{equation}
The scattering matrix is then given by
\begin{align}
  \mathbf{S}(\omega) &= \mathbf{I} - i\mathbf{W}^{+}(\omega - \mathbf{H}_{\mathrm{\textrm{eff}}})^{-1} \mathbf{W}, \\
  \mathbf{H}_{\mathrm{\textrm{eff}}} &= \mathbf{H} - \frac{i}{2} \mathbf{W}\mathbf{W}^{+},
\end{align}
where $\omega$ is the frequency.  After sampling $\mathbf{S}$, we can generate the spectral transmission matrix $\mathbf{A}$, which has size $M\times N$ and is defined by $A_{mn} = |S_{m0}(\omega_n)|^2$, where $\omega_1,\dots,\omega_N$ are a set of equally-spaced discrete frequency channels.

Next, we apply these RMT samples to the spectra and reconstruction schemes described in Section~\ref{sec:sm_dataset}.  As noted in the main text, the reconstruction MSE can be decomposed into two terms, a noise-induced term and a bias term:
\begin{equation}
  \mathbb{E} \, \big\|\hat{\mathbf{x}}-\mathbf{x}\big\|_{2}^{2}
  \;\,\ge\;
  \sigma_\epsilon^{2}\,\mathrm{Tr}(\mathbf{G}^{+})
  \,+\, \big\| \, \mathbb{E}[\hat{\mathbf{x}}]-\mathbf{x} \, \big\|_{2}^{2}.
\end{equation}
We will analyze each term separately.

\textit{Noise-induced MSE}---In Fig.~\ref{fig.MSE_RMT_parameters}\textbf{a}--\textbf{c}, we present simulation results showing the noise-induced MSE (i.e., the difference in MSE with and without noise) versus $\operatorname{Tr}[\mathbf{G}^{+}]$, for different choices of RMT parameters.  Within each subplot, we draw different ensembles by varying the RMT coupling parameter $\gamma$ defined in Eq.~\eqref{Wdef}.  The different subplots show different choices of the $M$ (the number of measurement channels), with a fixed number of frequency channels $N = 100$.  In each of the subplots, (a) the under-determined regime $M < N$, (b) $M = N$, and (c) the over-determined regime $M > N$, we find that the noise-induced MSE for the pseudoinverse closely follows the prediction of $\sigma_\epsilon^2 \operatorname{Tr}[\mathbf{G}^{+}]$ derived from the Cram\'er--Rao lower bound.  The results for NN reconstruction follow a similar trend, with MSE increasing with $\operatorname{Tr}[\mathbf{G}^{+}]$. However, we observe deviations from linearity at higher values of $\operatorname{Tr}[\mathbf{G}^{+}]$, indicating a more complex relationship than is captured by the Cram\'er--Rao theory.

\begin{figure}  \includegraphics[width=\columnwidth]{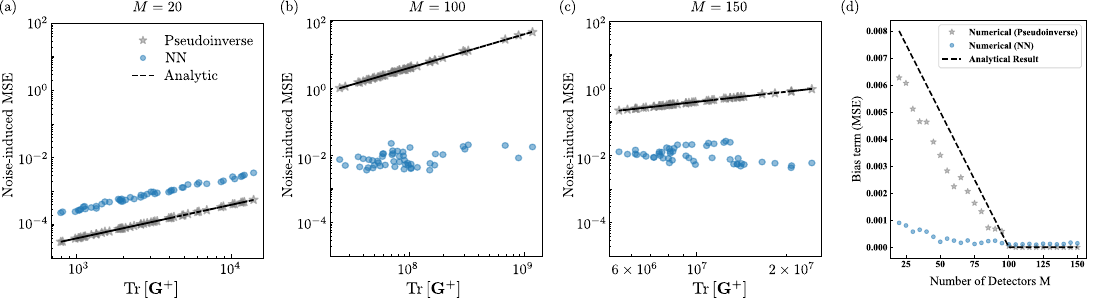}
  \caption{(a)--(c) Noise-induced MSE versus $\operatorname{Tr}[\mathbf{G}^{+}]$, obtained from RMT simulations with pseudoinverse and NN reconstruction schemes.  Each data point is an ensemble average over fixed RMT parameters; the different subplots correspond to different $M \in \{20, 100, 150\}$, with fixed $N = 100$, and in each subplot we vary $\gamma \in [0.005, 0.01]$.  The number of cavity modes is $P = 500$.  The dashes show the Cram\'er--Rao bound $\sigma_\epsilon^{2}\,\mathrm{Tr}(\mathbf{G}^{+})$, which quantitatively fits the pseudoinverse results.  (d) Bias term for different $M$, with $N = 100, \gamma = 0.005$ and all other parameters the same as in (a)--(c).  The pseudoinverse-based reconstruction experiences significantly increasing bias as $M$ is decreased below $N$; for NN reconstruction, the bias increases less sharply.
  }
  \label{fig.MSE_RMT_parameters}
\end{figure}

\textit{Bias term.}---Figure~\ref{fig.MSE_RMT_parameters}(d) plots the bias term, representing the intrinsic reconstruction error in the noise-free limit. This error arises solely from the projection of the signal $\mathbf{x}$ onto the $(N-M)$-dimensional null space of $\mathbf{A}$, which cannot be inverted linearly.
Mathematically, the bias for the pseudoinverse estimator is given by $\text{Tr}[(\mathbf{I} - \mathbf{A}^+ \mathbf{A}) \mathbf{R}_{xx}]$, where $\mathbf{R}_{xx} = \mathbb{E}[\mathbf{x}\mathbf{x}^T]$ is the correlation matrix of the input spectra. For a random scattering matrix, the null space orientation is isotropic. Consequently, the bias is determined by the trace of the correlation matrix (total energy) scaled by the dimensionality reduction ratio:
\begin{equation}
\text{MSE}_{\text{bias}} \approx \frac{N - M}{N} \cdot \frac{\operatorname{Tr}(\mathbf{R}_{xx})}{N} = \left(1 - \frac{M}{N}\right) \langle x^2 \rangle.
\end{equation}
Conversely, in the over-determined regime ($M \ge N$), the null space vanishes, and the bias term becomes strictly zero. The analytical prediction (dashed line), calculated using the empirical second moment $\langle x^2 \rangle \approx 0.01$, captures this behavior perfectly, including the linear descent and the sharp transition to zero at $M=100$. Notably, the Neural Network reconstruction (blue circles) yields a significantly lower bias than the linear baseline in the under-determined region, indicating that the model successfully leverages learned spectral priors to recover information within the null space.

\section{Details of spectrometer resolution derivation}
\label{SM:RMT_theory}

Here, we present further details about the derivation leading to the results, given in the main text, for $\mathbb{E}[\operatorname{Tr}(\mathbf{G}^{+})]$ and the effective spectrometer resolution.

First, we discuss the approximation leading up to Eq.~(7) in the main text, which simplifies the normalization factor in the correlation coefficient by a replacement
\begin{equation*}
  \|\mathbf{a}_i\|\|\mathbf{a}_j\| \rightarrow (\cdots)\;T_0^2/M.
\end{equation*}
In the strong scattering regime characterized by the GOE or GUE ensembles, the complex scattering coefficients $S_{m0}$ follow a circular complex Gaussian distribution \cite{lehmann1995chaotic, fyodorov1997statistics}. Consequently, the transmission intensities $A_{mn} = |S_{m0}(\omega_n)|^2$ follow a negative exponential (Rayleigh) distribution:
\begin{equation}
P(A) = \frac{1}{\langle A_{mn} \rangle} \exp\left(-\frac{A}{\langle A_{mn} \rangle}\right),
\end{equation}
where $\langle A_{mn} \rangle = T_0/M$ is the mean transmittance of a single detection channel ($T_0 = M\langle A_{mn}\rangle$ denotes the total transmittance, as defined in the main text).  This distribution has the property that the standard deviation equals the mean: $\sigma_A = \langle A_{mn} \rangle$.  This justifies the approximation we used in the main text; specifically, the $L_2$ norm of the zero-mean fluctuation vector $\mathbf{a}_j$ scales with the standard deviation as
\begin{align}
  \|\mathbf{a}_j\| &\approx \sqrt{M}\sigma_A = \sqrt{M}\,\langle A_{mn} \rangle = \frac{T_0}{\sqrt{M}} \\
  \Rightarrow \;\;\;
  \|\mathbf{a}_i\|\|\mathbf{a}_j\| &\approx M \langle A_{mn} \rangle^2 = \frac{T_0^2}{M}.
\end{align}
Since $C_{ij}$ is the inner product of two vectors of length $M$, the normalized correlation is $\mathrm{corr}(\mathbf{A}_i, \mathbf{A}_j) \approx M C_{ij}/T_0^2$, as stated in Eq.~(7) of the main text.

Next, we discuss why the spectral correlation function has a Lorentzian form, as stated in Eq.~(8) of the main text.  We work in the Ericson regime (strongly overlapping resonances), i.e., $\Gamma\gg\Delta$ where $\Delta$ is the mean level spacing, and $\Gamma$ is the mean resonance width.  In this regime, spectral averages are equivalent to ensemble averages due to the property of self-averaging (ergodicity) \cite{ericson1960fluctuations, weidenmuller2009random}. The Weisskopf-Wigner result of RMT states that the survival (dwell-time) distribution of modes in the cavity is exponential:
\begin{equation}
P(t)=\frac{1}{\tau_d}\,e^{-t/\tau_d},\qquad
\Gamma_{\rm W}\equiv \frac{1}{\tau_d}=
\frac{\Delta}{2\pi}\sum_{c=1}^{M} T_c.
\label{eq:A_Weisskopf}
\end{equation}
Here, $T_c$ are the channel transmission coefficients.

Let $E_m(E)$ denote the complex field on detector pixel $m$. The first-order (mutual) field correlation for an energy offset $\epsilon$ is the Fourier transform of $P(t)$:
\begin{align}
G_1^{(m)}(\epsilon) &\equiv 
\big\langle E_m(E)\,E_m^{*}(E+\epsilon)\big\rangle \\
&=\int_{0}^{\infty}\! P(t)\,e^{\,i\epsilon t}\,dt \\
&=\frac{1}{1-i\,\epsilon \tau_d}. 
\label{eq:A_G1}
\end{align}
Hence, the normalized intensity correlation function is a Lorentzian:
\begin{equation}
\big|G_1^{(m)}(\epsilon)\big|^{2}
=\frac{1}{1+(\epsilon \tau_d)^2}
=\frac{\Gamma_{\rm corr}^{\,2}}{\epsilon^{2}+\Gamma_{\rm corr}^{\,2}},
\end{equation}
where
\begin{equation}
\qquad \Gamma_{\rm corr}\equiv 1/\tau_d=\Gamma_{\rm W}.
\label{eq:A_Lorentz_field}
\end{equation}

\section{Correlation Length for Under-determined Systems}
\label{sec:sm_underdetermined}

Here, we derive the modified scaling function $\tilde{J}(a)$ given in Eq.~(15) of the main text.  This function is used in the under-determined regime ($M < N$) of the reconstructive spectrometer.  For later convenience, let us define
\begin{equation}
  \beta = M/N < 1.
\end{equation}
In this regime, the Gram matrix $\mathbf{G} = \mathbf{A}^T\mathbf{A}$ can be viewed as a Toeplitz matrix generated by a $f(\theta)$ that has been modified by the finite observation window.

Mathematically, the entries of $\mathbf{G}$ are formed by the autocorrelation of the transmission amplitudes over a finite window of size $M$. In the spectral domain (the domain of $f$), this finite windowing corresponds to a convolution of the asymptotic Lorentzian symbol $f_{\textrm{Lorentz}}(\theta)$ with the Fej\'er kernel $F_M(\theta)$:
\begin{equation}
  \tilde{f}(\theta) = (f_{\textrm{Lorentz}} * F_M)(\theta) = \int_{-\pi}^{\pi} f_{\textrm{Lorentz}}(\phi) F_M(\theta - \phi) d\phi,
\end{equation}
where
\begin{equation}
  F_M(\theta) = \frac{1}{M} \frac{\sin^2(M\theta/2)}{\sin^2(\theta/2)}
\end{equation}
is the spectral representation of the Bartlett (triangular) lag window.

\begin{figure}[t]
  \centering
  \includegraphics[width=0.8\columnwidth]{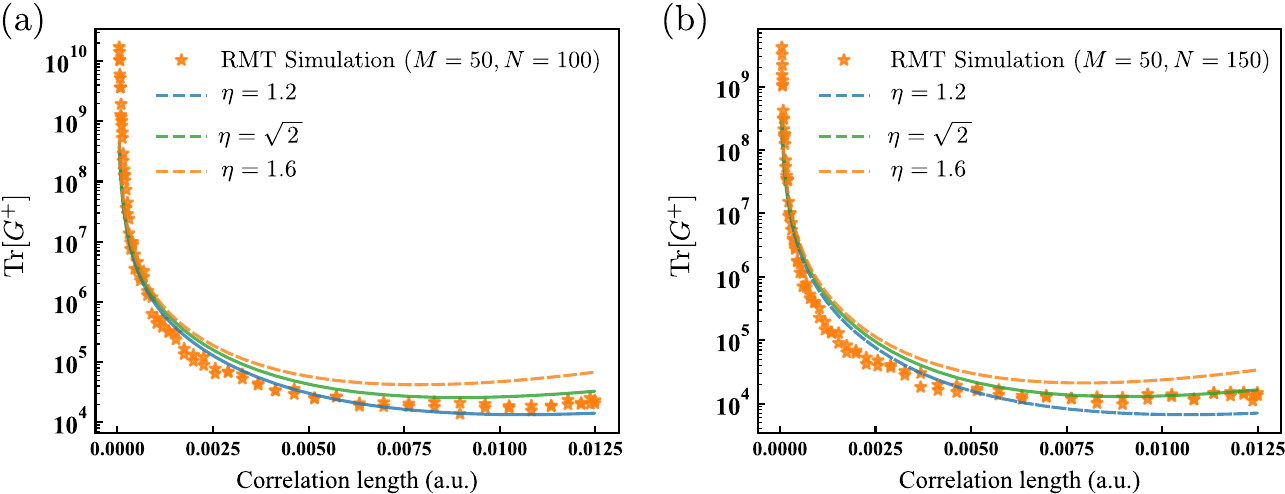}
  \caption{Sensitivity of the theoretical prediction for $\operatorname{Tr}[\mathbf{G}^{+}]$ to the broadening factor $\eta$.  Solid curves show the analytical predictions with $\eta = 1.2$ (blue), $\sqrt{2}$ (orange), and $1.6$ (green); markers denote RMT simulation results.  (a) $M = 50$, $N = 100$ ($\beta = 0.5$).  (b) $M = 50$, $N = 150$ ($\beta = 1/3$).  The parameter settings are identical to those in Fig.~2(c) of the main text.  All three values of $\eta$ yield qualitatively similar curves, but $\eta = \sqrt{2}$ provides the best overall fit across both configurations.}
  \label{fig.eta_sensitivity}
\end{figure}

Let us note that $\operatorname{Tr}(\mathbf{G}^+)$ is dominated by the behavior of $\tilde{f}(\theta)$ near its global minimum (the spectral gap).  The convolution with the Fej\'er kernel has two competing effects on the effective spectral correlation length $a_{\textrm{eff}}$ that governs this minimum:

\begin{itemize}
\item \textit{Linear Geometric Scaling}---The reduction in the number of independent measurement channels from $N$ to $M$ compresses the frequency axis relative to the integration time. To first order, this corresponds to a rescaling of the correlation parameter $a \to \beta a$.

\item \textit{Spectral Broadening}---The convolution with the Fej\'er kernel (which has a main lobe width $\sim 4\pi/M$) smears out the sharp features of the Lorentzian symbol. This smoothing lifts the minimum of the symbol and broadens the spectral valley, increasing the correlation length relative to pure linear scaling.
\end{itemize}

To capture both effects analytically, we introduce the following ansatz for the effective correlation length $a_{\textrm{eff}}$ in the modified trace formula:
\begin{equation}
  a_{\textrm{eff}} = \eta \beta a.
\end{equation}
Here, $\beta$ accounts for the geometric reduction of dimensionality, and $\eta > 1$ is a shape correction factor accounting for the Fejér broadening.  Let us now refer to the definition of the $J(a)$ function in Eq.~(11) of the main text, reproduced here for convenience:
\begin{equation}
  J(a) \equiv \frac{2\sinh(\pi a)}{\pi^2 a^{2}}
  \tan^{-1}\!\Bigl[\tanh\bigl(\tfrac{\pi a}{2}\bigr)\Bigr].
  \label{Jadef}
\end{equation}
We seek a form
\begin{equation}
  \tilde{J}(a) \approx \frac{2\sinh(\pi a_{\textrm{eff}})}{\pi^2 a^2}.
\end{equation}
By comparing to the RMT ensemble in the limit of strong correlations ($a \gg 1$), we find that the broadening factor converges to $\eta \approx \sqrt{2}$, i.e., the Bartlett windowing broadens the effective spectral correlation width by a factor of $\sqrt{2}$ relative to a hard rectangular truncation.  We note that the results are not sensitive to the precise value of $\eta$: values in the range $\eta \in [1.2, 1.6]$ all give reasonable agreement with the RMT data, with $\eta = \sqrt{2} \approx 1.414$ providing the best overall fit (see Fig.~\ref{fig.eta_sensitivity}).  This value remains consistent across ensembles with different numbers of internal modes ($P = 200$--$1000$) and different $M/N$ ratios ($\beta = 0.2$--$0.9$).  Thus, we set $a_{\textrm{eff}} = \sqrt{2} \frac{M}{N} a$, and substituting this into \eqref{Jadef} yields
\begin{equation}
  \tilde{J}(a) = \frac{2\sinh\left(\sqrt{2}\frac{M}{N}\pi a\right)}{\pi^2 a^{2}} \tan^{-1}\left[\tanh\left(\frac{\pi a}{2}\right)\right].
\end{equation}
The pre-factor $1/(\pi^2 a^2)$ and the $\tan^{-1}$ term remain dependent on the original $a$ as they relate to the normalization of the total energy and the bulk spectrum, which are less sensitive to the edge smoothing effects than the exponential tail governed by the $\sinh$ term.

\section{The Properties of the Inverse Trace Formula}
\label{sec:sm_trace}

Here, we discuss the $J(a)$ function, which plays a central role in our results for the spectrometer's performance.  In the limit where the spectral correlation length is much smaller than the sampling interval ($\Gamma_{corr} \ll \Delta \omega$), the system acts as a spectrometer with independent, uncorrelated channels (the "white" channel limit). Applying the small-angle approximations $\sinh(x) \approx x$ and $\tan^{-1}(\tanh(x)) \approx x$, we find:
\begin{equation}
    \lim_{a \to 0} J(a) \approx \frac{2(\pi a)}{\pi^2 a^{2}} \cdot \left(\frac{\pi a}{2}\right) = 1.
\end{equation}
Thus, we obtain:
\begin{equation}
  \mathbb{E}[\operatorname{Tr}(\mathbf{G}^{+})]
  \propto \frac{M^2 N}{|M-N|}\frac{1}{T_0^2}.
\end{equation}
This recovers the fundamental noise floor for a reconstructive spectrometer with orthogonal transmission columns (i.e., uncorrelated measurements), consistent with standard RMT predictions for Wishart matrices \cite{tulino2004random}.  In this case, the reconstruction error is mainly dependent on the mean transmittance $T_0$. The spectral correlation length $\Gamma_{corr}$ does not contribute to the error, as the channels are effectively independent. 

Next, let us consider the highly correlated regime ($a \gtrsim 1$). This corresponds to the spectral correlation length far exceeding the channel spacing ($\Gamma_{corr} \gg \Delta \omega$), meaning that the transmission matrix is ill-conditioned (i.e., adjacent columns become nearly collinear). Using the approximation $\sinh(\pi a) \approx \frac{1}{2}e^{\pi a}$ and noting that $\tanh(\frac{\pi a}{2}) \to 1$ (implying $\tan^{-1}(1) = \pi/4$), we obtain
\begin{equation}
    J(a) \approx \frac{e^{\pi a}}{\pi^2 a^2} \cdot \frac{\pi}{4} = \frac{e^{\pi a}}{4\pi a^2}.
\end{equation}
This exponential growth in $J(a)$ indicates that the reconstruction error diverges exponentially for highly correlated systems if the mean transmittance $T_0$ remains constant. However, in chaotic scattering systems, $T_0$ and $a$ are physically coupled; increasing the coupling strength typically enhances $T_0$, which can offset the penalty from $J(a)$. The optimal design point emerges from this competition, as seen in Fig.~2(c) of the main text; it is the point where the gain in signal throughput ($T_0^2$) outweighs the exponential penalty from spectral correlations ($J(a)$).

\begin{figure}[t]
  \centering
  \includegraphics[width=0.45\columnwidth]{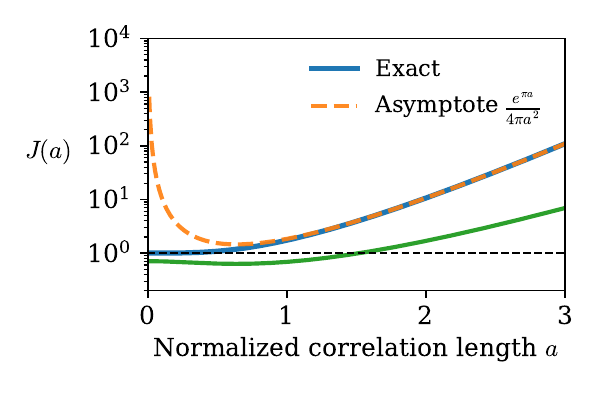}
  \caption{ 
  Variation of the scaling function $J(a)$ with the normalized correlation length $a = \Gamma_{\text{corr}}/\Delta \omega$. The solid blue line represents the exact analytical form for the over-determined case ($M \ge N$), which converges to the asymptotic exponential behavior $\sim e^{\pi a}$ (orange dashed line) for large correlations. The green solid line depicts the modified function $\tilde{J}(a)$ for an under-determined system ($M/N = 0.5$), exhibiting suppressed error growth due to spectral smoothing. All functions converge to $1$ at $a=0$, representing the baseline error for uncorrelated channels. }
  \label{fig.J_a_analysis}
\end{figure}

In the under-determined case ($M < N$), the finite observation window modifies the effective correlation length. As derived in Sec.~\ref{SM:RMT_theory}, the scaling function $\tilde{J}(a)$ incorporates a factor of $\sqrt{2}\beta$ (where $\beta = M/N < 1$) inside the hyperbolic sine term in Eq.~(15) of the main text.

Mathematically, this factor effectively reduces the rate of exponential growth compared to the over-determined case. Physically, this robustness arises from the spectral smoothing effect of the Fejér kernel (Bartlett windowing). By convolving the spectral density with the finite window, the deep spectral valleys (which correspond to near-zero eigenvalues) are broadened and lifted. This prevents the smallest eigenvalues from vanishing as rapidly as in the untruncated case, thereby suppressing the divergence of the pseudoinverse trace.

Fig.~\ref{fig.J_a_analysis} illustrates the behavior of $J(a)$ and $\tilde{J}(a)$ across the full range of $a$, highlighting the universal baseline at $a=0$ and the divergent behavior at large $a$.

\section{Super-resolution Regime}
\label{sec:sm_superres}

\begin{figure*}[t]
\centering
\includegraphics[width=\textwidth]{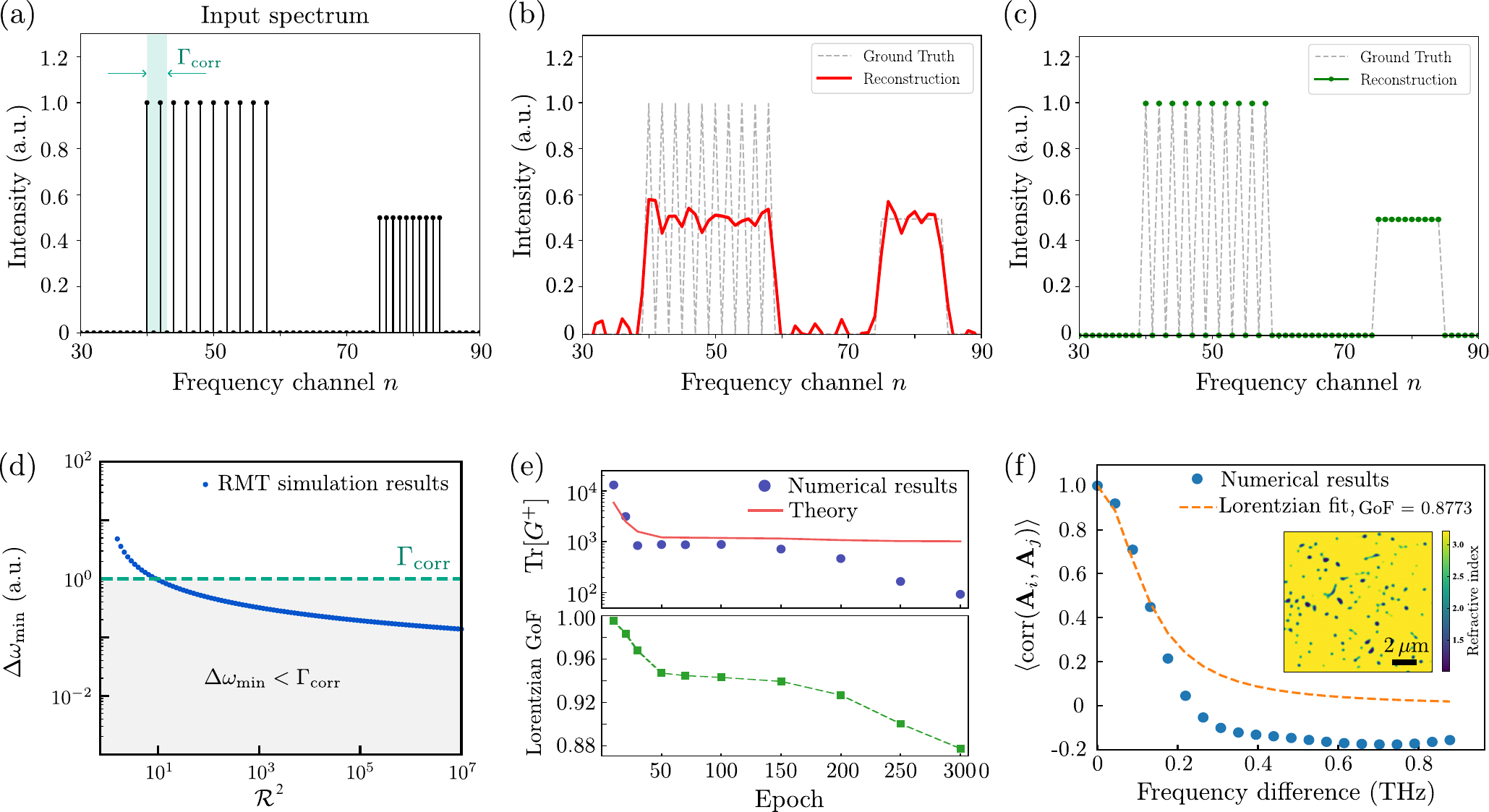}
\caption{(a) An input spectrum (black lines) consisting of $N = 100$ frequency channels spaced by $\Delta \omega$, with $\Gamma_{\text{corr}} = 3 \Delta \omega$ (green shaded region), implying strong spectral overlap.  The number of detectors if $M = 150$. (b) Reconstruction result at $\mathcal{R} \approx 10$ (red line).  The solver fails to resolve the individual comb lines (gray dashes). (c) Reconstruction result at $\mathcal{R} \approx 3\times 10^3$ using the same scatterer, with the comb lines now well-resolved (green circles). (d) Effective resolution $\Delta \omega_{\min}$ (blue line) versus $\mathcal{R}^2$, with $M = 150$ and $N = 100$. As $\mathcal{R}$ increases, the resolution drops below $\Gamma_{\text{corr}}$ (red dashes), entering the super-resolution regime (gray shaded area). (e) Top panel: variation of $\operatorname{Tr}[\mathbf{G}^{+}]$ (blue circles) and the corresponding theoretical prediction (red curve) versus number of optimization epochs, as a numerical inverse design scheme is employed to minimize $\operatorname{Tr}[\mathbf{G}^{+}]$ in a full-wave FEM simulation. Bottom panel: the corresponding Lorentzian goodness-of-fit, defined as the Pearson correlation coefficient between the actual spectral correlation data and an ideal Lorentzian function with $\Gamma_{\textrm{corr}}$ as the FWHM. (f) Spectral correlation function of the final optimized structure (blue circles) and the Lorentzian fit (orange dashes). Inset: the optimized refractive index distribution.}
\label{fig:super_resolution}
\end{figure*}

In the main text, we defined the effective spectral resolution $\Delta \omega_{\min}$ as the minimum frequency separation at which two spectral lines can be resolved with a reconstruction error below a threshold $\mathcal{E}_{\mathrm{th}}$. In this section, we apply this definition to analyze the super-resolution regime $\Delta \omega_{\min} < \Gamma_{\mathrm{corr}}$.

Conventionally, the resolution of a speckle-based spectrometer is assumed to be limited by the spectral correlation width $\Gamma_{\text{corr}}$, analogous to the Rayleigh criterion in imaging. This heuristic states that spectral features separated by less than $\Gamma_{\text{corr}}$ (the frequency scale over which speckles are highly correlated) cannot be distinguished. However, this perspective overlooks the critical role of the signal-to-noise ratio. In the context of computational spectral reconstruction, ``resolution'' is not a hard physical barrier but a parameter estimation problem. Two spectral features separated by $\Delta \omega < \Gamma_{\text{corr}}$ can be resolved if the signal-to-noise ratio is sufficiently high to distinguish their slightly different ``fingerprints''.

We illustrate this principle in Fig.~\ref{fig:super_resolution}(a)--(c). We consider an input spectrum consisting of discrete frequency comb lines separated by a spacing $\Delta \omega$. The transmission matrix $A$ is simulated using the RMT model (see Sec.~\ref{sec:RMT}) with a correlation length significantly larger than the frequency channel spacing ($\Gamma_{\text{corr}} = 3 \Delta \omega$). As shown in Fig.~\ref{fig:super_resolution}(a), the individual channels are spectrally overlapped within the correlation width. Using the same transmission matrix, we perform spectral reconstruction at two different noise levels, corresponding to different values of the effective SNR $\mathcal{R}$, defined in Eq.~(14) of the main text. When the reconstruction is performed at $\mathcal{R} \approx 10$ [Fig.~\ref{fig:super_resolution}(b)], the algorithm fails to resolve individual lines, recovering only the broad envelope of the spectrum. The reconstruction error is high, and the effective resolution is coarse, consistent with the classical $\Gamma_{\text{corr}}$ limit. At $\mathcal{R} \approx 3\times 10^3$ [Fig.~\ref{fig:super_resolution}(c)], however, the comb lines are resolved even though the scatterer remains identical. Evidently, sufficiently high $\mathcal{R}$ allows the spectrometer to operate in the super-resolution regime.

Figure~\ref{fig:super_resolution}(d) quantifies this relationship. Based on the RMT simulations, we plot the theoretical resolution limit $\Delta \omega_{\min}$ as a function of $\mathcal{R}^2$ by varying the noise level $\sigma_{\epsilon}$ from $10^{-5}$ to $10^{-2}$ for 100 samples. The other system parameters are fixed at $M=150$, $N=100$, and $\delta^2_{th} = 10^{-3}$. The dashed red line indicates the physical correlation limit $\Gamma_{\text{corr}}$, which is normalized to 1. As the SNR increases, $\Delta \omega_{\min}$ decreases monotonically, eventually crossing below the $\Gamma_{\text{corr}}$ threshold to enter the super-resolution regime.

\section{Generalization to Poisson noise}
\label{sec:sm_poisson}

The main text assumes additive Gaussian noise with constant variance $\sigma_\epsilon^2$.  Here we show that the key scaling laws carry over to the Poisson (shot-noise) regime.

For shot-noise-limited detection, $y_m \sim \mathrm{Poisson}\!\bigl((\mathbf{A}\mathbf{x})_m\bigr)$, and the Fisher information matrix is
\begin{equation}
  G_{ij}^{(\mathrm{P})} = \sum_{m=1}^{M} \frac{A_{mi}\,A_{mj}}{(\mathbf{A}\mathbf{x})_m}
  \;=\; \bigl[\mathbf{A}^T \mathrm{diag}(1/\mathbf{A}\mathbf{x})\,\mathbf{A}\bigr]_{ij}.
  \label{eq:poisson_fim}
\end{equation}
Unlike the Gaussian case, $\mathbf{G}^{(\mathrm{P})}$ depends on the input spectrum~$\mathbf{x}$.  For a broadband (approximately flat) input, $(\mathbf{A}\mathbf{x})_m \approx \bar{y} \propto T_0$, so
\begin{equation}
  \mathbf{G}^{(\mathrm{P})} \approx \bar{y}^{\,-1}\,\mathbf{A}^T\mathbf{A}
  \;=\; \bar{y}^{\,-1}\,\mathbf{G}.
\end{equation}
The Cram\'er--Rao bound therefore becomes $\mathrm{MSE} \ge \bar{y}\;\mathrm{Tr}[\mathbf{G}^{+}]$, and the reconstruction error scales as
\begin{equation}
  \mathrm{MSE}_{\mathrm{Poisson}} \propto \frac{1}{T_0}\, J(a),
\end{equation}
compared to $\mathrm{MSE}_{\mathrm{Gaussian}} \propto \sigma_\epsilon^2 / T_0^2\, J(a)$.  The scaling function $J(a)$, and hence the roles of $\Gamma_{\mathrm{corr}}$, the conditions for super-resolution, and the resolution formula, all remain identical.  The only difference is that the transmittance penalty weakens from $T_0^{-2}$ to $T_0^{-1}$, implying that in the shot-noise-limited regime, thicker and more strongly scattering media become relatively more favorable.

\section{Inverse design study}
\label{sec:sm_inverse}

As noted in the conclusion of our paper, it may be possible for individual extremal scatterers to surpass the performance bounds predicted by our theory, due to certain assumptions embedded in the theory.  In particular, we have assumed that the scatterer obeys conventional speckle statistics with Lorentzian spectral correlations, which holds empirically for many real disordered scatterers and can be derived from RMT  \cite{ericson1960fluctuations, lehmann1995chaotic, fyodorov1997statistics, weidenmuller2009random}, but is \textit{not} immutable.

To study this issue, we conduct a full-wave FEM simulation study in which the structure is inverse-designed using the adjoint optimization method \cite{Molesky2018, Kroker2024}, to minimize the variance bound $\operatorname{Tr}[\mathbf{G}^{+}]$ \cite{yu2025wavelength}.  In Fig.~\ref{fig:super_resolution}(e), the upper panel shows how $\operatorname{Tr}[\mathbf{G}^{+}]$ varies with the number of optimization epochs (blue circles), along with the corresponding theoretical value (red curve). The theory successfully tracks the actual decrease in $\operatorname{Tr}[\mathbf{G}^{+}]$ for the first 100 or so epochs, but begins to fail after $\sim 150$ epochs.  In the lower panel of Fig.~\ref{fig:super_resolution}(e), we see that this failure coincides with a decrease in the goodness-of-fit to a Lorentzian spectral correlation function.

In Fig.~\ref{fig:super_resolution}(f), we plot the correlation function for the final optimized structure (whose refractive index is shown inset), which is indeed significantly non-Lorentzian.  It thus appears that the $\operatorname{Tr}[\mathbf{G}^{+}]$ minimization process works by introducing non-Lorentzian spectral correlations, moving the system away from the ``fully developed speckle'' regime, or equivalently, altering the dwell-time distribution of the cavity eigenmodes.

\begin{figure}[t]
  \centering
  \includegraphics[width=0.5\columnwidth]{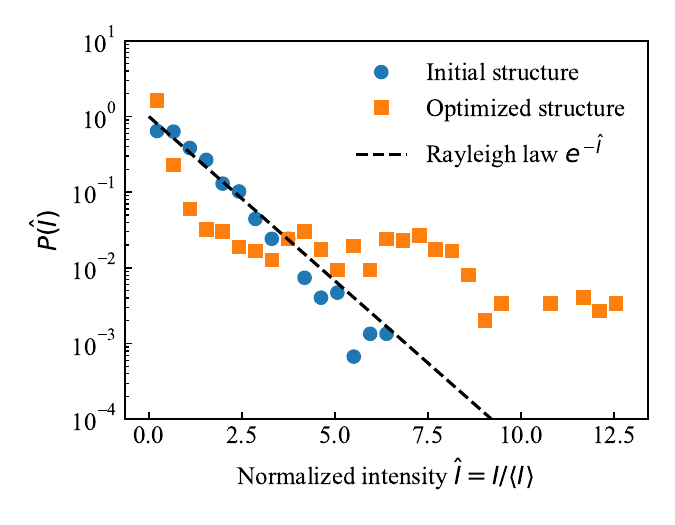}
  \caption{Intensity statistics of the initial and inverse-design-optimized structures.  Symbols show the probability density function (PDF) of the normalized transmitted intensity $\hat{I} = I/\langle I \rangle$, obtained from the FDTD transmission matrices by pooling all detection channels (each normalized by its own mean transmittance) over several early epochs (initial structure, blue circles) and late epochs (optimized structure, orange squares).  The dashed line is the Rayleigh law $P(\hat{I}) = e^{-\hat{I}}$ of fully developed speckle.  The optimized structure exhibits super-Rayleigh statistics, with second-moment ratio $\langle I^2 \rangle / \langle I \rangle^2 \approx 5.2$, compared to $\approx 1.7$ for the initial structure (Rayleigh value: 2).}
  \label{fig:speckle_pdf}
\end{figure}

To further characterize the optimized structure, Fig.~\ref{fig:speckle_pdf} shows the PDF of the normalized transmitted intensity $\hat{I} = I/\langle I \rangle$.  The initial structure closely follows the Rayleigh (negative-exponential) law characteristic of fully developed speckle, with second-moment ratio $\langle I^2 \rangle / \langle I \rangle^2 \approx 1.7$, close to the Rayleigh value of 2.  The optimized structure instead exhibits strongly \textit{super-Rayleigh} statistics: the PDF acquires a heavy tail extending beyond $\hat{I} \approx 10$, with $\langle I^2 \rangle / \langle I \rangle^2 \approx 5.2$.  Evidently, the optimization does not homogenize the transmission; instead, it concentrates the transmission into sparse, high-contrast spectro-spatial features on a darker background, which act as more distinguishable fingerprints for neighboring frequency channels.  This departure from Rayleigh statistics is consistent with the non-Lorentzian spectral correlations seen in Fig.~\ref{fig:super_resolution}(f), and confirms that the optimized scatterer has exited the fully-developed-speckle (Ericson) regime.

\clearpage

\end{widetext}

\end{document}